\documentclass[11pt]{article}
\usepackage{moriond,epsfig}

\def\Journal#1#2#3#4{{#1} {\bf #2}, #3 (#4)}

\def\PLB{{\em Phys. Lett.}  B}

\def\be{\begin{equation}}
\def\ee{\end{equation}}
\def\bea{\begin{eqnarray}}
\def\eea{\end{eqnarray}}

\newcommand{\met}{\ensuremath{{E}\!\!\!{/}_{T}} }
\begin{document}
\vspace*{4cm}
\title{SEARCHES FOR THE STANDARD MODEL HIGGS AT THE TEVATRON}

\author{P.M. JONSSON\\ (For the CDF and D\O\ Collaborations)}

\address{Imperial College London, Prince Consort Road,\\
London SW7 2BW, United Kingdom}

\maketitle\abstracts{{\bf Abstract.} Recent preliminary results obtained by the CDF and D\O\ collaborations on searches for the Standard Model (SM) Higgs boson in $p\bar{p}$ collisions at $\sqrt{s}$ = 1.96 TeV at the Fermilab Tevatron are discussed. The data, corresponding to integrated luminosities between $260 - 950 $ pb$^{-1}$, show no excess of a signal above the expected background in any of the decay channels examined. Instead, upper limits at 95\% Confidence Level (C.L.) on the cross section are established. For the first time, a combined SM cross section limit based on 14 orthogonal analysis channels from D\O\ is presented.}

\section{Introduction}
The search for the SM Higgs boson is one of the main challenges for particle physics and as such a high priority for the upgraded CDF and D\O\ detectors at Run II of the Tevatron. The Higgs boson, which is needed to explain the mechanism of electroweak symmetry breaking, is the only particle predicted in the SM which has not been discovered. Indirect constraints on the mass ($m_{H}$), which is a free parameter, from fits to the global set of electroweak data, favor a light SM Higgs boson of 89 GeV with an upper limit of 175 GeV at 95\% C.L~\cite{lepewwg}. The direct searches at LEP have already excluded a SM Higgs mass below 114.4 GeV~\cite{lep}. A discovery of the Higgs may be within reach by the end of Run II, thanks to the excellent accelerator and detector performance. The expected combined sensitivity to exclude a SM Higgs at 114 GeV at the Tevatron starts around $2$ fb$^{-1}$~\cite{tevwg} and both CDF and D\O\ have now each recorded an integrated luminosity of over $1$ fb$^{-1}$.

\section{Low Mass Searches, $m_{H}< 135$ GeV}
Higgs boson production cross sections in the SM are small at Tevatron energies, of the order of 1-0.1 pb depending on the production mechanism. Gluon fusion, $gg\rightarrow H$, is the dominant production mechanism. However, for masses below 135 GeV, where $H\rightarrow b\bar{b}$ decays dominate, the QCD background is overwhelming. The smaller but cleaner channels of associated $ZH$ and $WH$ production, with the vector bosons decaying into leptons can instead be used for direct searches. These analyses rely on efficient $b$-tagging and lepton identification as well as precise Monte Carlo (MC) modeling of the backgrounds. Progress is being made on understanding the $W/Z$ + jets background. A recent study from D\O\, for example, shows good detector level agreement up to $n_{jet} = 4$ between SHERPA 1.0.6, a matrix element + parton shower MC generator, and $950$ pb$^{-1}$ of selected $Z(\rightarrow e^{+}e^{-})+n_{jet}$ data~\cite{sherpa}.  

\subsection{$ZH\rightarrow \nu\bar{\nu}b \bar{b}$}
This channel has a good sensitivity because of the large $Z\rightarrow \nu \bar{\nu}$ and $H\rightarrow b \bar{b}$ branching ratios. Since the two $b$-jets are boosted, the final state contains a signature of acoplanar jets in contrast to typical QCD dijets. The main backgrounds are $W/Z$+jets, $WZ$, $ZZ$ and $t\bar{t}$ where the lepton or jets escape. Multijets, with mis-measured jets dominate the difficult instrumental backgrounds. CDF has performed a search with $289 $ pb$^{-1}$ of data~\cite{cdfzh} and D\O\ has analyzed $261 $pb$^{-1}$ of data~\cite{makoto}\footnote{In addition, this analysis is also used at D\O\ in the search for $WH\rightarrow l \nu b \bar{b}$ with a missed lepton to improve the $WH$ sensitivity.}. The event selections include 1 or 2 $b$-tagged jets and large missing transverse energy ($\met$). Since no significant excess is observed over the expected backgrounds, upper limits at 95\% C.L. are calculated for $\sigma_{ZH}\times BR(H\rightarrow b\bar{b})$. The resulting limits are shown, together with the other Tevatron limits, in Fig.~\ref{fig:comb}.

\subsection{$WH\rightarrow l \nu b \bar{b}$}
\begin{figure}[h]

\begin{center}
\epsfig{figure=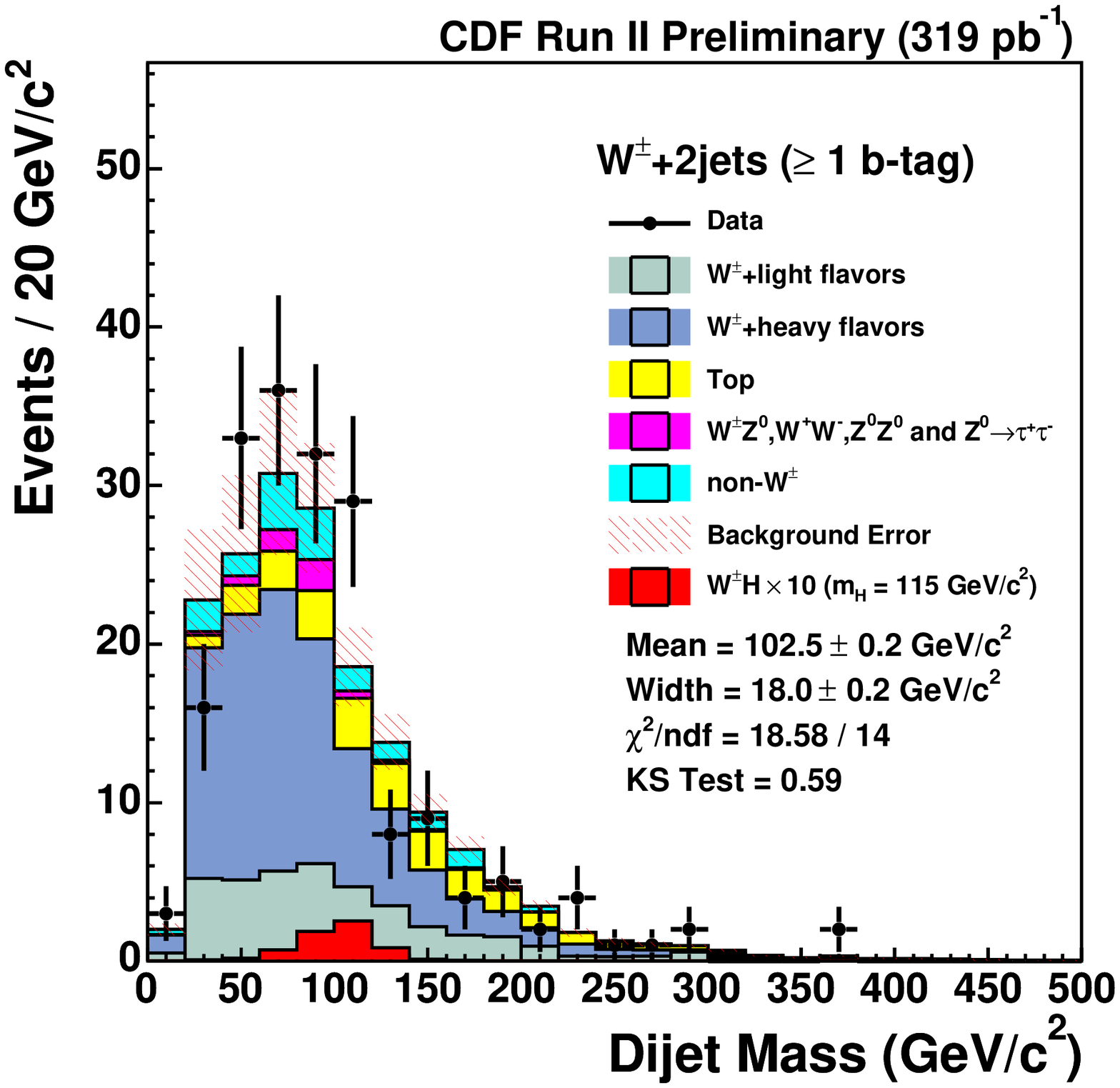,height=2.0in}
\epsfig{figure=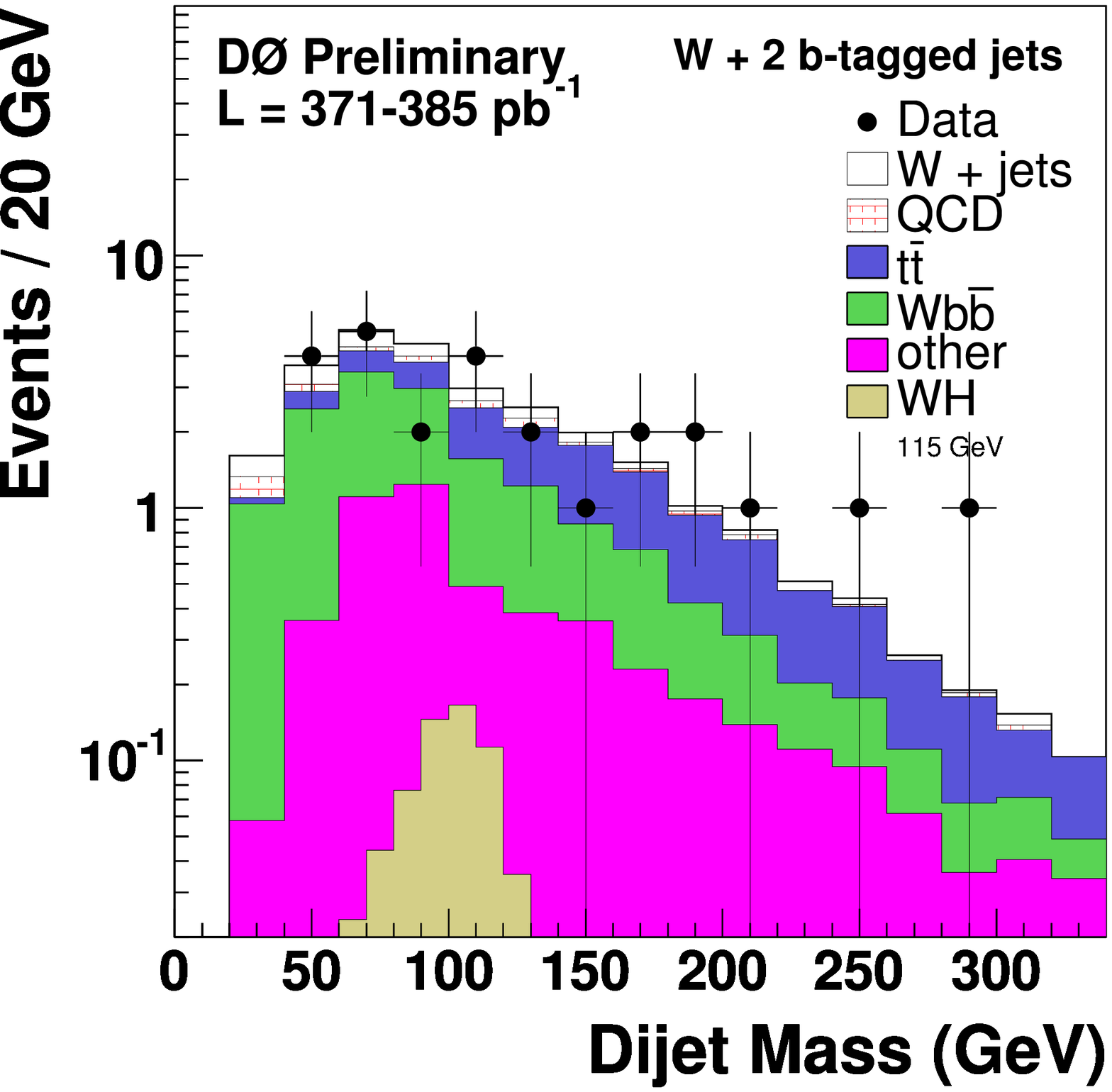,height=2.0in}
\caption{The dijet mass distribution in data from the $WH$ search along with the background expectations. Left: CDF. Right: The final double $b$-tagged events from D\O\ in logarithmic scale.}
\label{fig:WH}
\end{center}
\end{figure}

Associated $WH$ production, with the $W$ decaying to a charged lepton and a neutrino, provides perhaps the most promising channel. CDF has performed a search with $319$ pb$^{-1}$ of data~\cite{cdfwh} and D\O\ reports new preliminary results for both the electron and muon channel based on $378$ pb$^{-1}$ of data~\cite{d0wh}. The analysis selects candidate events by identifying the $W$ through its decay to an electron or muon and a neutrino. The events are required to have a central, isolated electron or muon with $p_{T} > 20$ GeV, $\met > 20$ GeV (D\O\ : $\met > 25$ GeV) and two jets with $E_{T} > 15$ GeV  in a pseudo rapidity region of $  | \eta | < 2.0$ ( D\O\ : $E_{T} > 20$ GeV within $ | \eta | < 2.5$ ).
In addition, one or two $b$-tagged jets are required. Fig.~\ref{fig:WH} shows the invariant dijet mass distribution of the final event samples from CDF and D\O\ . No significant excess is observed and therefore upper limits at 95\% C.L. are calculated. The limits are illustrated in Fig.~\ref{fig:comb}. 

\section{High Mass Searches, $m_{H} > 135$ GeV}
At higher Higgs masses, where they are kinematically possible, $H\rightarrow WW^{(*)}$ decays with subsequent electronic and/or muonic decays of the $W$s, provide promising search channels with manageable backgrounds.

\subsection{ $WH\rightarrow WWW^{(*)}$}
Because of the like-signed leptons in the final state, much of the SM backgrounds from diboson and $t\bar{t}$ production can be reduced in this channel. 

D\O\ has performed a recent search with $363-384$ pb$^{-1}$ of data~\cite{d0www} (CDF has a previous result based on the analysis of $194 $ pb$^{-1}$ of data~\cite{cdfwww}). The analysis selects two isolated like signed leptons ($ee, e\mu$ or $\mu\mu$) with $p_{T} > 15$ GeV and $\met > 20$ GeV. After final event selection 6 events remain in data, which is in agreement with the predicted SM background. The resulting upper limit on the cross section is illustrated in Fig.~\ref{fig:comb}.

\subsection{ $H\rightarrow WW^{(*)}$}
The signature of this signal consists of two isolated, oppositely charged leptons together with large $\met$. Direct reconstruction of the Higgs mass is not possible, due to the two neutrinos in the final state. Instead, the spin-correlations between the decay products of the Higgs boson can be used to suppress the background. The charged leptons from the signal events tend to be collinear. CDF has analyzed $360$ pb$^{-1}$ of data~\cite{cdfww} and D\O\ has presented a new result based on $950$ pb$^{-1}$ for the $e\mu$ and $ee$ channels~\cite{d0ww}. After final event selection, which also includes a veto around the $Z$ mass, good agreement with the SM background predictions is observed. The resulting CDF cross section limit, obtained from fits to the $\Delta \phi_{ll}$ distribution, is illustrated together with the D\O\ result in Fig.~\ref{fig:comb}.

\section{Combined Limits}
\begin{figure}[h]
\begin{center}
\epsfig{figure=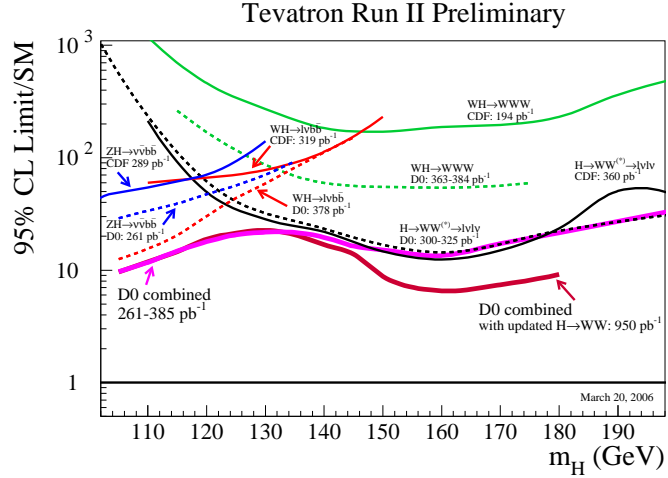,height=2.5in}
\caption{The ratio of the excluded cross section (at 95\% C.L.) to the SM cross section as a function of the Higgs mass for the various Higgs searches at the Tevatron. The effect of including the new $H\rightarrow WW^{(*)}$ analysis in the combined D\O\ limit is clear. At $m_{H} = 160$ GeV the new limit from D\O\ is only a factor 7 from the SM.}
\label{fig:comb}
\end{center}
\end{figure}

 D\O\ has, for the first time, calculated a combined upper SM Higgs cross section limit at 95\% C.L. based on fourteen orthogonal search channels\footnote{The single and double $b$-tagged events from the $ZH$ and $WH$ searches are treated as separate channels.}~\cite{d0comb}. The resulting combined limit is shown together with the other SM Higgs limits from the Tevatron in Fig.~\ref{fig:comb}. The SM cross section is still a factor 15(7) away from the new upper limit at $m_{H} = 115(160)$ GeV. 

\section{Conclusions}
The new preliminary results presented at this conference together with the recent performance of the Tevatron and the experiments, are very encouraging for the Higgs searches at Run II. The combined preliminary limit on the SM cross section from D\O\ demonstrates a large improvement in sensitivity over previous results. With upcoming improvements to the analyses and combination of more channels from both experiments, the Tevatron sensitivity can be expected to approach the SM level for a $115$ GeV Higgs with 2 fb$^{-1}$ of data.
   
\section*{Acknowledgments}
I would like to thank my colleagues from the CDF and D\O\ collaborations for providing material for this talk and the organizers of Rencontres de Moriond for a most stimulating conference.

\end{document}